\documentclass{PoS}
\usepackage{amsmath,amssymb, bm, cite}

\title{Signs for the onset of gluon saturation in exclusive photo-production of vector mesons}

\ShortTitle{Signs of gluon saturation at the LHC}

\author{\speaker{Martin Hentschinski}\thanks{Collaboration with Alfredo Arroyo
    Garcia is gratefully acknowledged.}\\
        Departamento de Actuaria, F\'isica y Matem\'aticas
Universidad de las Am\'ericas Puebla
Ex-Hacienda Santa Catarina Martir S/N,
San Andr\'es Cholula
72820 Puebla, Mexico\\
        E-mail: \email{martin.hentschinski@udlap.mx}}

\author{Krzysztof Kutak\\
       The H. Niewodnicza\'nski Institute of Nuclear Physics,  Polish Academy of Sciences,ul. Radzikowskiego 152, 31-342, Cracow, Poland \\
        E-mail: \email{krzysztof.kutak@ifj.edu.pl}}

\abstract{We investigate the energy dependence of the photo-production
  cross-section of vector mesons $J/\Psi$ and $\Upsilon$,  measured by both HERA experiments H1 and ZEUS in
  electron-proton collisions and by LHC experiments ALICE, CMS and
  LHCb in ultra-peripheral proton-proton and ultra-peripheral
  proton-lead collisions. Our study uses 2 particular fits of
  inclusive unintegrated gluon distribution, based on non-linear
  Balitsky-Kovchegov evolution (Kutak-Sapeta gluon; KS) and
  next-to-leading order Balitsky-Fadin-Kuraev-Lipatov evolution
  (Hentschinski-Sabio Vera-Salas gluon; HSS). We find that linear
  next-to-leading order BFKL evolution can only describe  production
  at highest energies, if perturbative corrections are increased to
  unnaturally large values; rendering this corrections small, the
  growth with energy is too strong in the LHC region and the
  description of $J/\Psi$ data fails. For the KS gluon we find that an
  accurate description of  $J/\Psi$ data is possible if non-linear
  corrections to low x QCD evolution are taken into account; without
  such correction a description of data fails. We interpret this
  observation as a clear signal for the presence of high gluon
  densities in low x the proton, characteristic for the onset of gluon
  saturation.} 

\FullConference{7th Annual Conference on Large Hadron Collider Physics - LHCP2019\\
		20-25 May, 2019\\
		Puebla, Mexico}

\begin{document}

\section{Introduction}

The power-like rise of the gluon distribution at small $x$, with
$x = M^2/s$ the ratio of the hard scale $M$ and the center of mass
energy $\sqrt{s}$, is well established both experimentally
\cite{Abramowicz:2015mha} and theoretically
\cite{BFKL1,Hentschinski:2012kr}. Nevertheless such a power-like rise
cannot continue forever. Bounds imposed by unitarity require that the
observed rise in $x$ will eventually  slow down and come to hold. From a
theory point of view one expects the formation of an over occupied
system of gluons, which eventually leads to saturation of gluon
densities \cite{Gribov:1984tu}; finding convincing phenomenological
evidence for gluon saturation is still one of the open problems of
Quantum Chromodynamics (QCD). The evolution from the low to large
gluon densities is described by a set of nonlinear evolution
equations, known as
Balitsky-Jalilian-Marian-Iancu-McLerran-Weigert-Leonidov-Kovner; its
frequently used mean field version is given by the Balitsky Kovchegov
(BK) \cite{Balitsky:1995ub} evolution equation. A suitable process to
study the low $x$ gluon at the Large Hadron Collider is provided by
exclusive photo-production of vector mesons. For this observable a
large amount of data has been collected both for the production of
$J/\Psi$ and $\Upsilon$ vector mesons. The hard scale is in both cases
provided by the heavy quark mass, {\it i.e.} the charm ($J/\Psi$) and
bottom ( $\Upsilon$) quark mass. In the case of the $J/\Psi$ one is
therefore able to reach very small $x$ values at a low transverse
scale, which allows for the potential observation of saturation
effects. Photo-production of the $\Upsilon$ provides on the other hand
a cross-check of the description well in the perturbative domain,
where such effects are genuinely expected to be absent.

\section{Methodology}
\label{sec:methodology}

To establish the presence of effects related to the presence of high
gluon densities and the on-set of gluon saturation, it is necessary
but not sufficient to establish that frameworks which include
corrections due to gluon saturation are capable to describe data.
Instead it is necessary to establish the break-down of frameworks
which do not include such effects. While the established description
of QCD cross-sections in the perturbative low-density regime is based
on collinear factorization and DGLAP evolution, we argue that it does
not provide suitable benchmark for this particular observable: the
$x$-dependence of collinear parton distribution functions is obtained
from a fit to data at an initial hard scale which is usually of the order of the  charm mass. As a consequence no
perturbative evolution takes place for the description of $J/\Psi$
data. The description of the $x$ dependence is therefore a pure
fit. The  perturbative QCD enters only  through
DGLAP evolution, {\it i.e.} through the evolution from $J/\Psi$ to the
$\Upsilon$ scale. For such evolution the following observations apply:  a) $\Upsilon$
data are placed at large $x$ values than corresponding $J/\Psi$ data
b) DGLAP evolution shifts large $x$ input to lower $x$ {\it i.e.} the
low $x$ $\Upsilon$ depends on the fit to $J/\Psi$ data at intermediate
$x$, where absence of saturation effects is firmly established c)
higher twist effects associated with high gluon densities die away
fast within collinear DGLAP evolution. It is therefore far more useful to compare to linear
Balitsky-Fadin-Kuraev-Lipatov (BFKL) low $x$ evolution instead: Perturbative
QCD is being used to determine the low $x$ behavior of the gluon,
based on fits of an unintegrated gluon distribution at intermediate
values of $x$. A breakdown of such an description and the simultaneous
description of data by a framework which incorporates high density
effects, constitutes then a sign for the presence of effects related
to gluon saturation. For a detailed discussion we refer the interested
reader to \cite{Garcia:2019tne}, see also \cite{Bautista:2016xnp}.

\section{Results}
\label{sec:results}

\begin{figure}[p!]
  \centering
  \includegraphics[width=.95\textwidth]{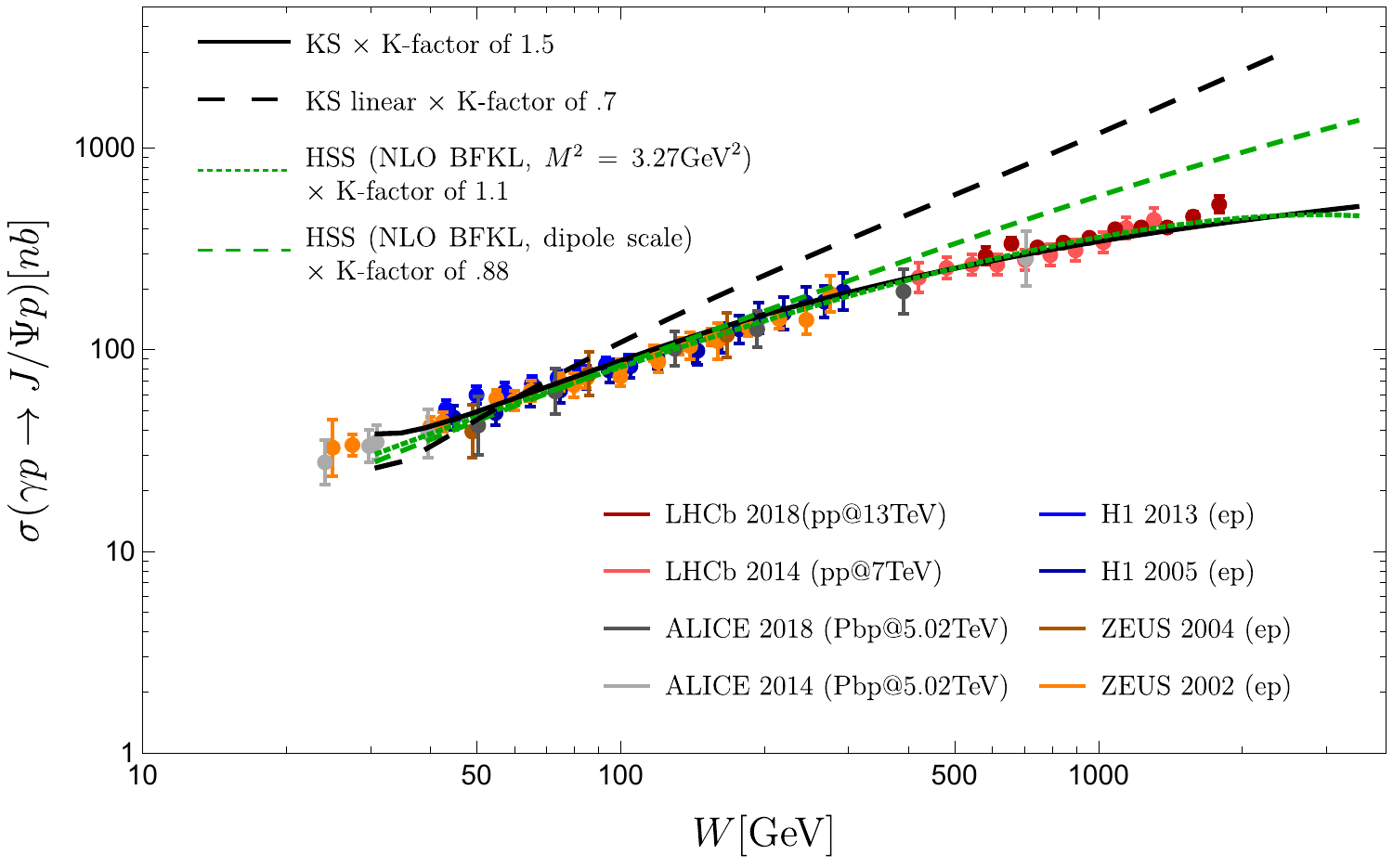}\\
\vspace{1cm}
  \includegraphics[width=.95\textwidth]{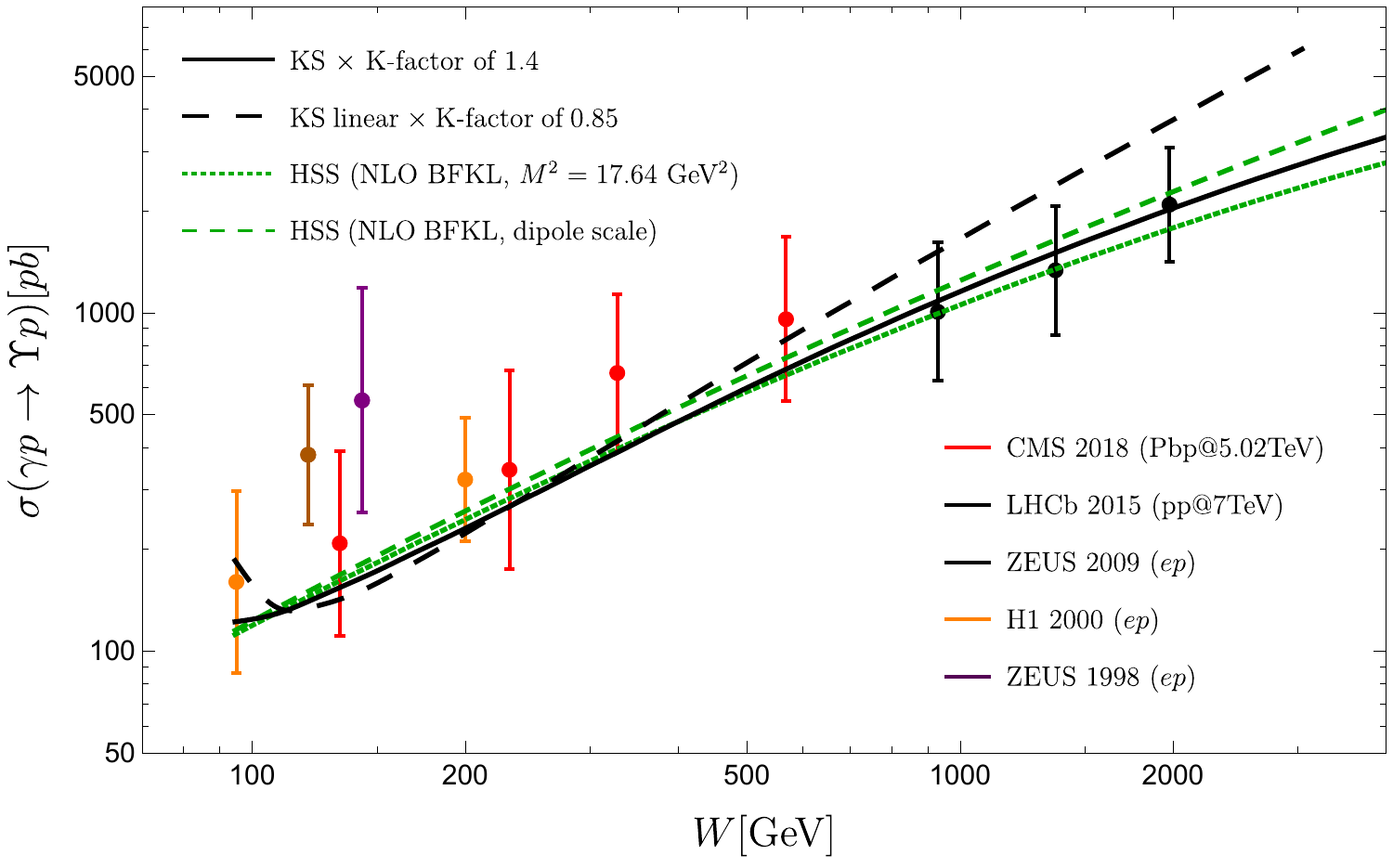}
  \caption{\it Energy dependence of the $J/\Psi$ and $\Upsilon$
    photo-production cross-section as provided by the KS and HSS gluon
    distribution. The HSS distribution with dipole size scale
    corresponds to a specific scale setting for the HSS gluon
    discussed in Sec.~\ref{sec:results}.  For the $J/\Psi$ we
    further display photo-production data measured at HERA by ZEUS and H1 collaborations
    \cite{Chekanov:2002xi} as well as LHC data obtained
    from ALICE and LHCb \cite{TheALICE:2014dwa}.  For the $\Upsilon$ cross-section we show
    HERA data measured by H1 and ZEUS \cite{Adloff:2000vm}  and LHC data by LHCb and  CMS \cite{Aaij:2015kea}.}
  \label{fig:results}
\end{figure}

In the following we will compare photo-production data to two
particular QCD fits to combined HERA data which are subject to QCD low
$x$ evolution.
As the perturbative benchmark we will use the Hentschinski-Salas-Sabio Vera
(HSS)\cite{Hentschinski:2012kr} unintegrated gluon which is based on
linear NLO BFKL evolution.  To assess the importance of non-linear
terms in low $x$ evolution equations, we use a particular solution to
BK-evolution, with initial conditions fitted to combined HERA data by
Kutak-Sapeta (KS) \cite{Kutak:2012rf}. Both HSS and KS gluon include
sub-leading terms related to the resummation of collinear logarithms. Our results are shown in
Fig.~\ref{fig:results}. 
The results of our study for the fixed renormalization scales ({\it i.e.} an  external scale related to the heavy quark mass) can be found in
 Fig.~\ref{fig:results}: continuous, black lines correspond to
 the KS-gluon and dotted, green lines to the HSS-gluon at fixed
 scales. At first sight it appears that both the HSS (linear)
 and the KS (non-linear) gluon describe the energy dependence of the
 data. One might therefore conclude that there is no need for
 non-linear evolution associated with the presence of high gluon
 densities and that saturation effects are absent. Taking however a
 closer look at the dipole cross-section associated with the HSS-gluon
 one realizes that it takes the following form (see
 \cite{Garcia:2019tne} for a detailed discussion). With the dipole cross-section $\sigma_{q\bar{q}}(x, r$ obtained from the unintegrated gluon density $\mathcal{F}(x, {\bm k}^2)$ through $ 
 \sigma_{q\bar{q}}(x, r) = \frac{4 \pi}{N_c} \int \frac{d^2 {\bm
                           k}}{{\bm k}^2}
\left(1-e^{i {\bm k}\cdot {\bm r}}\right) \alpha_s  {\cal F}(x, {\bm k}^2) \,
$, 
 one has 
\begin{align}
  \label{eq:1}
  \sigma_{q\bar{q}}^{(\text{HSS})} (x, r) & = \alpha_s   \hat{\sigma}_{q\bar{q}}^{(\text{HSS})} (x, r) ,
&
\hat{\sigma}_{q\bar{q}}^{(\text{HSS})} (x, r) & =  \hat{\sigma}_{q\bar{q}}^{(\text{dom.})} (x, r)  +   \hat{\sigma}_{q\bar{q}}^{(\text{corr.})} (x, r)  ,
\end{align}
where 
\begin{align}
  \label{eq:2}
\hat{\sigma}_{q\bar{q}}^{(\text{dom})} (x, r, M^2)   &= \int\limits_{\frac{1}{2} - i \infty} ^{\frac{1}{2} + i \infty}\!\! \frac{d \gamma}{2 \pi i} \left(\frac{4}{r^2 Q_0^2} \right)^\gamma \frac{ \bar{\alpha}_s(M \cdot Q_0)}{\bar{\alpha}_s(M^2)}
 f(\gamma, Q_0, \delta, r)   \left(\frac{1}{x}\right)^{\chi\left(\gamma, M^2 \right)}\notag \\
\hat{\sigma}_{q\bar{q}}^{(\text{corr.})} (x, r, M^2)   &= \int\limits_{\frac{1}{2} - i \infty} ^{\frac{1}{2} + i \infty}\!\! \frac{d \gamma}{2 \pi i} \left(\frac{4}{r^2 Q_0^2} \right)^\gamma \frac{ \bar{\alpha}_s(M \cdot Q_0)}{\bar{\alpha}_s(M^2)}
 f(\gamma, Q_0, \delta, r)   \left(\frac{1}{x}\right)^{\chi\left(\gamma, M^2 \right)}\notag \\
&\hspace{-2cm} \times   \frac{\bar{\alpha}_s^2\beta_0  \chi_0 \left(\gamma\right) }{8 N_c} \log{\left(\frac{1}{x}\right)}
  \Bigg[- \psi \left(\delta-\gamma\right)
 +  \log \frac{M^2r^2}{4} - \frac{1}{1-\gamma} - \psi(2-\gamma) - \psi(\gamma) \Bigg]\;,
\end{align}
\begin{figure}[t]
  \centering
 \parbox{.49\textwidth}{
 \includegraphics[width=.45\textwidth]{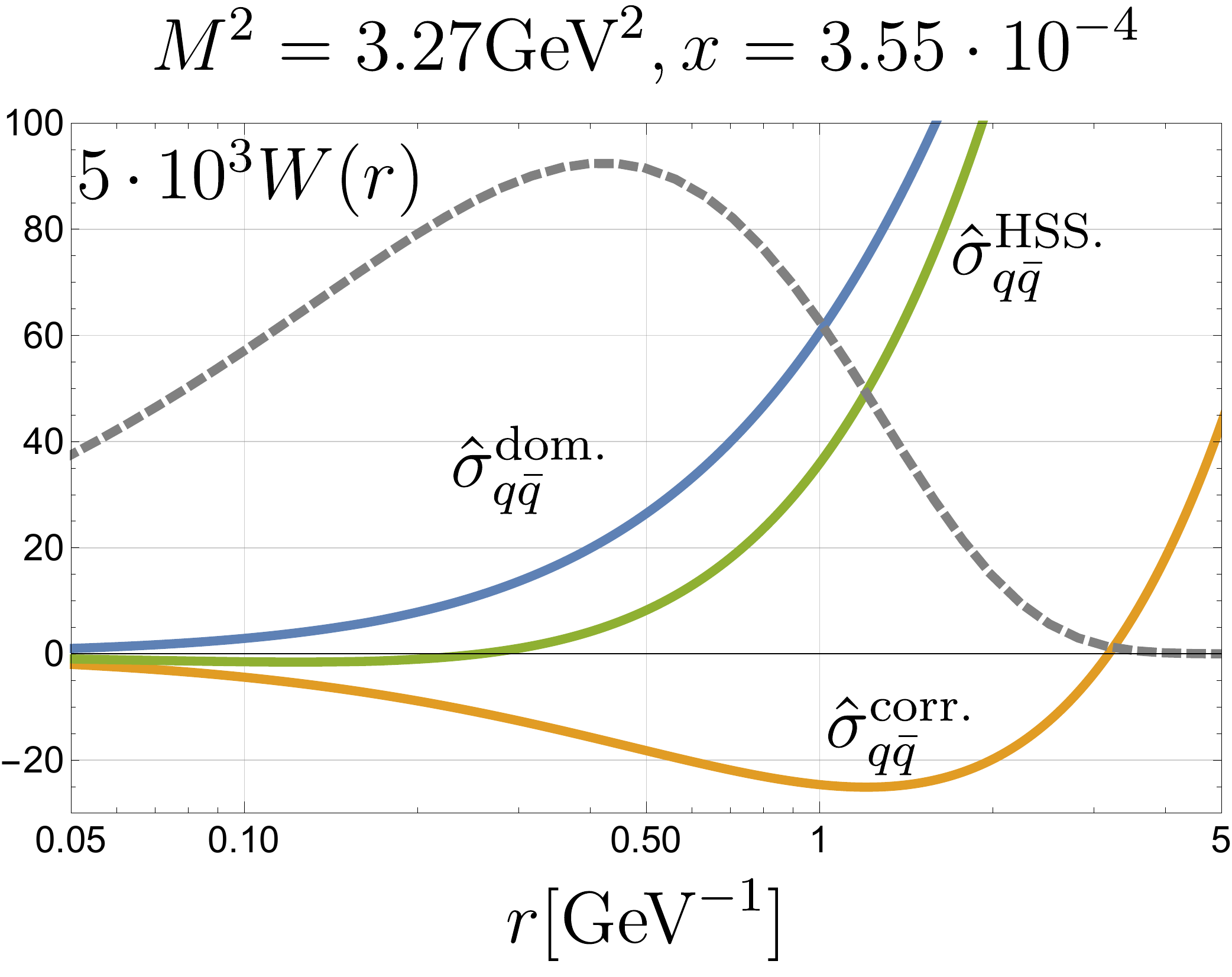}}
\parbox{.49\textwidth}{
  \includegraphics[width=.45\textwidth]{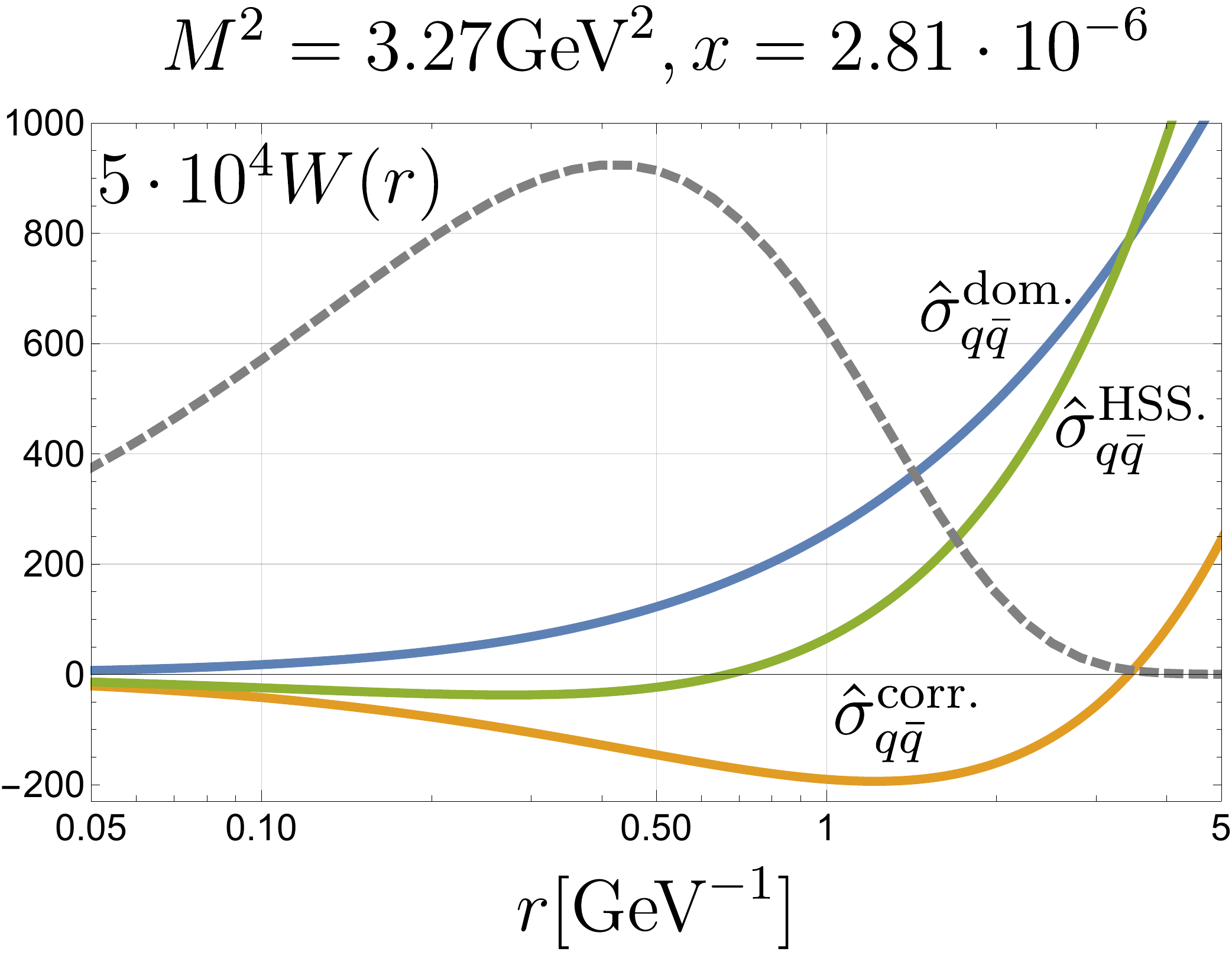}} \\
\parbox{.49\textwidth}{\center (a)}\parbox{.49\textwidth}{\center (b)}
 \parbox{.49\textwidth}{
   \includegraphics[width=.45\textwidth]{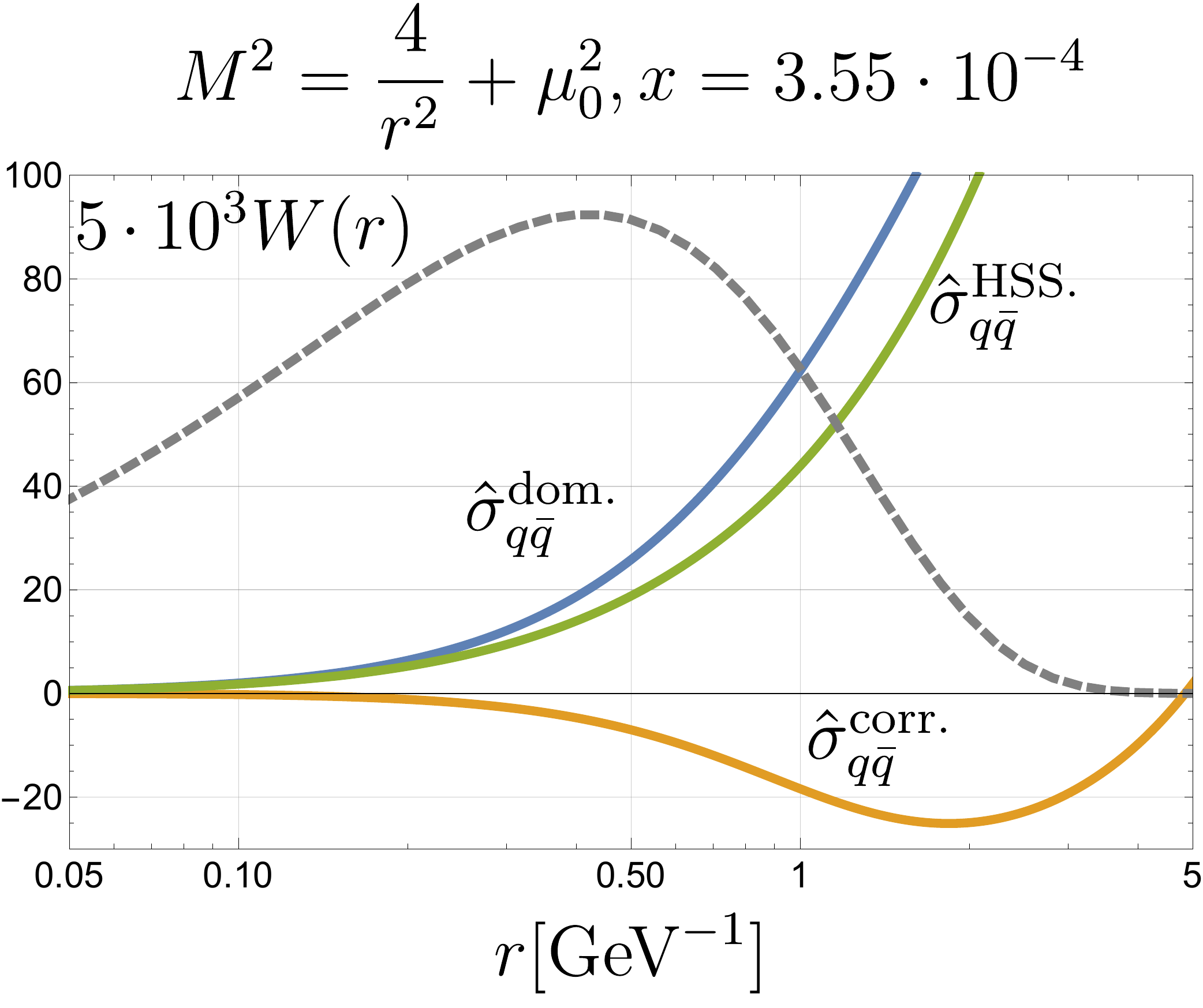} }
 \parbox{.49\textwidth}{
 \includegraphics[width=.45\textwidth]{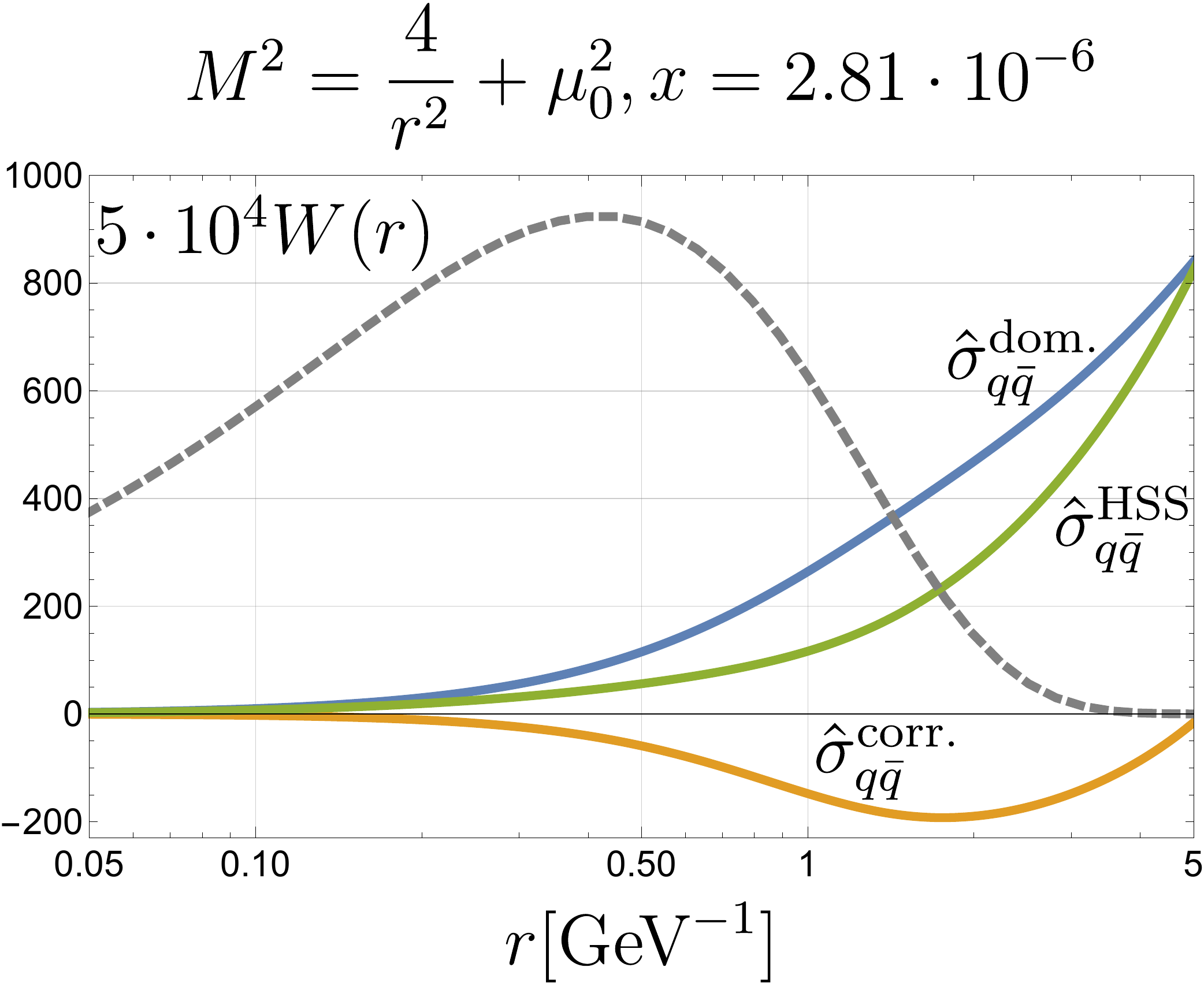}} \\
\parbox{.49\textwidth}{\center (c)}\parbox{.49\textwidth}{\center (d)}\\

  \caption{HSS dipole cross-section (with an overall factor of $\alpha_s$ extracted) at fixed ($J/\Psi$-scale, top row) and $r^2$-dependent scale (bottom row) in units of GeV$^{-2}$. The function   $W(r)$  indicates the typical dipole sized probed in $J/\Psi$ photo-production. }
  \label{fig:HSS_dipole}
\end{figure}
The second term constitutes a perturbative correction for the
next-to-leading logarithmic resummation, since it is of order
$\alpha^{n+1} \ln^n(1/x)$, $n \geq 0$. Even though this term is
formally sub-leading, it grows with decreasing $x$ and the logarithm
can eventually compensate smallness of the perturbative strong
coupling $\alpha_s$.  This is behavior is explored in
Fig.~\ref{fig:HSS_dipole}. We find that if we set the hard-scale to a
fixed value $M^2 = 3.27$~GeV$^2$, the perturbative correction is
indeed small for $x$ values typical for HERA kinematics, for which the
original fit has been performed. Turning on the other hand to the
smallest $x$ values probed in photo-production of $J/\Psi$s at the
LHC, the corrections supersedes the formally leading term; the
perturbative expansion is breaking down. There is an easy fix to this
problem, namely choosing an $r$-dependent renormalization scale. This
removes the $r$-dependent logarithm in the correction term and
consequently reduces the size of the correction in the critical
region. Having stabilized the linear NLO BFKL evolution, we return to
data: while the new solution (dashed green line) essential agrees
with the fixed scale solution for the $\Upsilon$ and for the $J/\Psi$ in the
HERA region (the observed deviation is of the order of a typical
variation of the renormalization scale, see \cite{Bautista:2016xnp}),
the stabilized linear NLO BFKL evolution overshoots data for the
$J/\Psi$ in the LHC region. At the same time the energy dependence of
the data is very well described by the non-linear KS gluon. To assess
the importance of the non-linearities in the solution, we also compare
to the KS-gluon with non-linearities turned off (dashed black line):
We observe that the linear KS-gluon overshoots data. We therefore
conclude that non-linear effects are essential to describe the energy
dependence of $J/\Psi$ data in the LHC region. We interpret this as a
clear sign for the onset of saturation effects in this region of phase
space.

\section{Conclusions}
\label{sec:conclusions}

The observed slow-down of the growth with energy is one of the core
predictions of gluon saturation.  To establish the observation made in
this letter it is therefore necessary to search for different
observables which probe the low $x$ gluon in a similar kinematic
regime and to increase further the theoretical accuracy of the
underlying framework. Steps to address the latter are currently
undertaken in \cite{vanHameren:2019ysa} (azimuthal de-correlations of
dijets) and \cite{Hentschinski:2017ayz,Hentschinski:2018rrf} (theory developments).

\end{document}